\documentclass[reprint,aps,prl,superscriptaddress,amsmath,amssymb,floatfix,footinbib,longbibliography]{revtex4-1}
\usepackage{graphicx}
\usepackage{dcolumn}
\usepackage{bm}% bold math
\usepackage{amsmath}
\usepackage{times}
\usepackage{color}
\usepackage[breaklinks=true,colorlinks,citecolor=blue,linkcolor=blue,urlcolor=blue]{hyperref}

\makeatletter

\newcommand{\Rmnum}[1]{\expandafter\@slowromancap\romannumeral #1@}
\makeatother

\begin{document}

\title{Anisotropic magnetization dynamics in Fe$_5$GeTe$_2$ at room temperature}
\author{Alapan Bera}
\affiliation{Department of Physics, Indian Institute of Technology Kanpur, Kanpur 208016, India}
\author{Nirmalya Jana}
\author{Amit Agarwal}
\author{Soumik Mukhopadhyay}
\email{soumikm@iitk.ac.in}
\affiliation{Department of Physics, Indian Institute of Technology Kanpur, Kanpur 208016, India}

\begin{abstract}
The determination of Lande's g-factor and the damping constant is central to extracting crucial information about the spin-spin and spin-orbit interactions in magnetically ordered systems. In insulating compounds based on 3d elements, the spin-orbit interaction can be effectively probed by investigating the ground state of the corresponding free 3d ion within the crystalline environment. For metallic systems, the problem is non-trivial because of additional band structure effects. Here, we investigate the anisotropic magnetization dynamics in bulk single crystalline Fe$_5$GeTe$_2$, a van der Waals 2D itinerant ferromagnet at room temperature, using broadband ferromagnetic resonance spectroscopy. Strikingly, contrary to the results of ab-initio calculations, we do not observe any intrinsic anisotropy of magnetization damping close to room temperature, suggesting diminished role of spin-orbit interaction. However, there is a sizable anisotropy in the Lande's g factor near room temperature which is attributed to anisotropic critical spin fluctuations.
\end{abstract}

\maketitle

{\it Introduction.---} 
Fe$_5$GeTe$_2$ (F5GT) is a recently discovered 2D van der Waals ferromagnet~\cite{AM1, AM2} with a notably elevated Curie temperature of $\mathrm{T_c \sim 310}$ K. It has garnered significant attention for its potential applications in spintronics~\cite{TANG20231, Yang, Alahmed, MTahir, DPalai}. The direct observation of topological spin textures, including (anti)skyrmions~\cite{Schmitt2022} (up to room temperature) and (anti)merons~\cite{gao, casas}, has sparked interest in both fundamental research and technological applications. These spin textures originate from spin-spin and spin-orbit interactions within the material, making it crucial to understand the magnetic characteristics of such systems. Despite the dominance of crystal field interaction over spin-orbit interaction for 3d ions like Fe$^{2+}$ and Fe$^{3+}$, there is a possibility that the orbital angular momentum may not be fully quenched in F5GT due to significant spin-orbit interaction. A recent first-principle study suggests a significant orbital contribution from non-equivalent iron sites in the magnetization of F5GT~\cite{Joe}, which influences its magnetization dynamics~\cite{Alahmed}. Thus, magnetization dynamics experiments are crucial
for investigating the role of anisotropy and spin-orbit interaction in magnetic systems. The spin-orbit contribution can be comprehensively probed using ferromagnetic resonance (FMR) spectra: (1) indirectly, by studying the angle dependence of the magnetization damping factor; and (2) directly, by studying the angle dependence of Lande's g-factor, the dimensionless form of the gyromagnetic ratio.

The magnetization damping determines the stability and lifetime of magnetic excitations in a solid. Both intrinsic (Gilbert damping) and extrinsic (induced by eddy currents, spin pumping, etc.) sources contribute to magnetization damping. The damping parameter is deduced from the resonant line-width in FMR experiments~\cite{Barman}. The FMR line-width is influenced by both intrinsic and extrinsic relaxation effects. In addition to intrinsic line-width broadening due to Gilbert damping, long-range fluctuations from variations in sample parameters such as internal fields, thickness, or crystal orientation can lead to frequency-independent inhomogeneous line broadening of the FMR signal. Short-range fluctuations due to two-magnon scattering can be suppressed when the magnetization is oriented perpendicular to the sample plane. Extrinsic contributions to line-width also arise from eddy currents and the field dragging effect between magnetically hard and easy directions. If all extrinsic contributions are neglected, the Gilbert damping parameter $\mathrm{\alpha}$ is expected to scale as $\mathrm{\alpha \sim D (E_F) |\Gamma^{-}|^{2}\tau}$~\cite{Li}, where $\mathrm{D(E_F)}$ is the density of states at the Fermi level $\mathrm{E_F}$, $|\Gamma^{-}|$ is the strength of the spin-orbit interaction, and $\mathrm{\tau}$ is the electron momentum scattering time. Generally, the Gilbert damping parameter is assumed to be an isotropic scalar, except in spin-orbit coupled systems with significant lattice or electronic structure anisotropy~\cite{Seib, Gilmore2010, Zhai, Kastani, Chen2018, Li, Tonig, Miranda}.

The spin-orbit interaction strength can be directly probed by measuring the g-factor from the resonant field in the FMR spectra, discussed in greater detail later. The \emph{almost} quenched ground state should still lead to the g-factor deviating marginally from the spin-only value of 2. This deviation from 2 reflects the presence of mixed-in $\mathrm{L > 0}$ states. However, we have to be cautious in drawing conclusions solely based on the deviation of the g-factor from its spin-only value. Close to the critical temperature, in addition to spin-orbit interaction, anisotropic critical fluctuations~\cite{Nagata1, Nagata2, Nagata3, Nagata4} can also lead to such deviations.

\begin{figure*}
    \centering
    \includegraphics[width=0.8\linewidth]{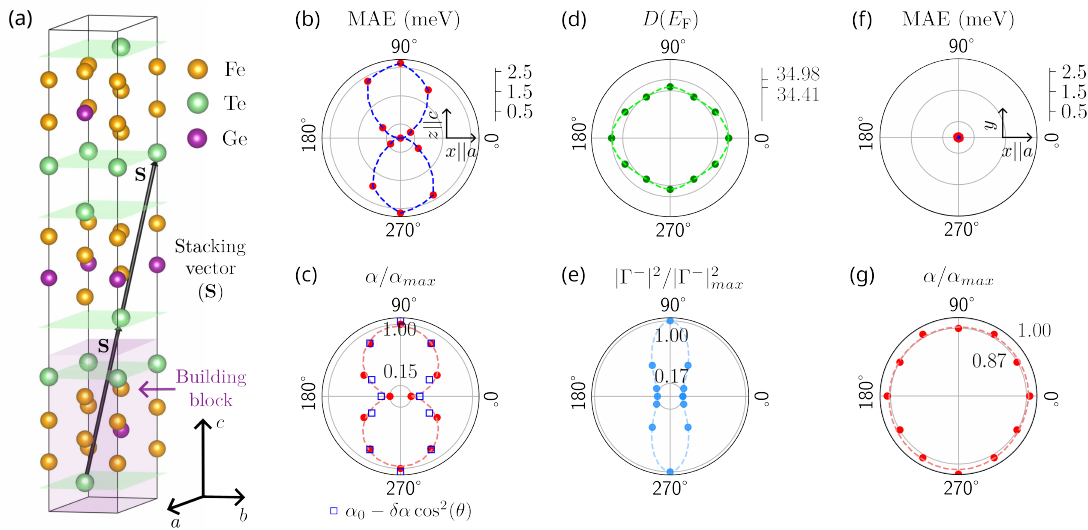}
    \caption{(a) Bulk unit cell consists of three building blocks (BB) of F5GT. The BBs are stacked by a vector $\textbf{S}$. 
    (b) Magnetic anisotropy energy (MAE) for different orientations of collinear spins by angle $\theta$ with the $x$-axis in the $x-z$ plane in presence of SOC. 
    (c) Variations of the Gilbert damping constant with the polar angle of the spin orientation,  plotted as red solid dots and extrapolated with red dashed lines. The large anisotropy in the damping constant ratio is two-fold symmetric, and the calculated values fit well to a form $\alpha_0-\delta \alpha~\cos^{2}\theta$, with $\alpha_0 \approx 1$ and $\delta \alpha \approx 0.8$. 
    (d) The angular variations of the density of states (DOS) at Fermi energy with the polar angle of the magnetization vector in the $x-z$ plane. The DOS has almost negligible variation with the orientation of the magnetization. 
    (e) The impact of the SOC on the damping anisotropy in terms of  $|\Gamma^{-}|^{2}$, on varying the polar angle of magnetization in the $x-z$ plane. This establishes that SOC is the dominant factor in inducing the giant anisotropy in the damping constant in our zero temperature calculations.  
    The (f) MAE and (g) the Gilbert damping constant do not vary much with the azimuthal angle of the magnetization vector in the $x-y$ plane.}
    \label{fig.1}
\end{figure*}

In this Letter, we provide clear evidence of magnetic anisotropy in ferromagnetic van der Waals single crystal Fe$_5$GeTe$_2$ (F5GT) at room temperature. This anisotropy arises from the interplay between anisotropic spin-spin interactions and enhanced magnetic fluctuations. The significant polar angle-dependent anisotropy of Lande's g-factor indicates the presence of anisotropic critical fluctuation induced magnetic anisotropy. Additionally, our observations show no intrinsic angular variation in the damping factor at room temperature, further confirming that orbital magnetism plays a negligible role in the g-factor anisotropy in F5GT.

{\it Ab initio calculations.---} Bulk F5GT crystallizes in the non-centrosymmetric space group R3m (number 160) \cite{stahl_18}. The unit cell [shown in Fig.~\ref{fig.1}a] comprises of three blocks with a stacking vector $\textbf{S} = (-a/3, b/3, c/3) = (2a/3, b/3, c/3)$, where $a, b, c$ are the three lattice parameters. Inside a block, Fe and Ge layers are sandwiched between two $a-b$ planes of Te atoms (green planes in Fig.~\ref{fig.1}a). 
The two nearest blocks are connected with van der Walls interaction due to large distance ($\sim$ 3.06 {\AA}) between two consecutive Te planes. The Fe atoms inside each block have finite magnetization. The exchange couplings between Fe atoms inside a block or between two blocks are ferromagnetic (FM).  
The collinear spins are aligned along the [100] crystal direction in presence of spin-orbit coupling (SOC). The FM F5GT is metallic with contributions of many bands at the Fermi energy, $E_{\mathrm{F}}$, as shown in SM-Fig.~1b in Sec. I of the supplementary material (SM)~\cite{SM}. As the spins rotate, the metallic Fermi surfaces also change.
The damping of magnetization dynamics depends on variations in the Fermi surface when rotating the spins. Changes in the magnetization orientation induce a non-equilibrium population imbalance of itinerant electrons, which relax to the equilibrium distribution over a scattering timescale ($\tau$)~\cite{Kambersky1970, Kambersky1984, Korenman1972, Kune2002}. The lagged excited states repopulate the empty equilibrium states by transferring angular momentum to the lattice, resulting in damping \cite{Gilmore2010, Kambersky1976, Gilmore2007}.

The matrix elements of the magnetization damping tensor are given by \cite{Miranda} 
\begin{equation}
    \alpha^{\mu \nu} = \frac{\mathrm{g} \pi \tau}{m \hbar} \int \frac{d\textbf{k}}{(2\pi)^{3}} \sum_{p, q} \eta(\epsilon_{\textbf{k}, p})\left(\frac{\partial \epsilon_{\textbf{k}, p}}{\partial \theta}\right)_{\mu}\left(\frac{\partial \epsilon_{\textbf{k}, q}}{\partial \theta}\right)_{\nu}~. 
    \label{eqn:spin_damping}
\end{equation}
Here, $p$ and $q$ ($\mu$ and $\nu$) represent the band (Cartesian coordinate) indices, and $\epsilon_{\textbf{k}, p}$ is the band dispersion at momentum $\textbf{k}$ for the $p$-th band. In Eq.~\eqref{eqn:spin_damping}, $\eta (\epsilon_{\textbf{k}, p}) = (\partial f(\epsilon)/\partial \epsilon)|_{\epsilon_{\textbf{k}, p}}$ with $f$ being the Fermi distribution function and $\tau$, $\mathrm{g}$, $m$ and $\hbar$ are the relaxation time, Land\'{e} g factor, magnetization and reduced Planck's constant, respectively. Here, $\frac{\partial \epsilon_{\textbf{k}, p}}{\partial \theta}$ captures the change in the band energy of the system on changing the direction of magnetization by a small amount. This change in band energy is primarily induced by the change in the spin-orbit coupling Hamiltonian of the system on changing the magnetization direction.  
For a collinear FM, assuming the local magnetization to be along the $z'$ direction, the scalar $\alpha$ is calculated by taking the average value, $\alpha = (\alpha^{x'x'} + \alpha^{y'y'})/2$~\cite{Miranda}.  

We present the variation of the calculated magnetic anisotropy energy (MAE) with polar angle (in the $x-z$ plane) in Fig.~\ref{fig.1}b.
The corresponding damping constant presented in Fig.~\ref{fig.1}c (normalized with its maximum value - $\alpha/\alpha_{max}$) shows a strong anisotropy in the $x-z$ plane, roughly following the MAE anisotropy. Our calculations indicate that $\alpha/\alpha_{max}$ is 6.67 times stronger along the $z$-axis than in any axis in the $x-y$ plane. Interestingly, we find that there is no anisotropy in either the MAE energy or in the magnetization damping term in the $x-y$ plane, as shown in Fig.~\ref{fig.1}f and \ref{fig.1}g for different magnetization directions. 
The variation in the MAE and the magnetization damping coefficient in the $x-z$ plane can arise either from the magnetization dependence of the density of states or from the interplay of spin-orbit coupling and the magnetization direction. To understand the mechanism of the large damping anisotropy of around $\sim$ 670\%, we present the variation of DOS at the Fermi energy [$D (E_{\mathrm{F}})$] with magnetization in Fig.~\ref{fig.1}d. We find that  $D (E_{\mathrm{F}})$ remains almost constant and shows no change with the magnetization direction with respect to the  $z-$axis. Thus, the damping anisotropy is likely to arise from the SOC. 

To probe the role of SOC in inducing the damping anisotropy further, we note that at low temperatures,
we can approximate the damping constant using the average of $\partial \epsilon_{\textbf{k}, p}$/$\partial \theta$ over the single particle states at $E_{\mathrm{F}}$. Using this approach, $\alpha$ is given by~\cite{Li, Fahnle_2008, Kambersky1970, Kambersky1976, Kambersky2007, Gilmore2007, Gilmore2010} 
\begin{equation}
    \alpha \sim D (E_{\mathrm{F}}) |\Gamma^{-}|^{2}\tau~.
    \label{eqn:dependence_of_alpha}
\end{equation}
Here, $|\Gamma^{-}|^{2} = (\partial \epsilon/\partial \theta)^{2}_{x} + (\partial \epsilon/\partial \theta)^{2}_{y}$ \cite{Gilmore2007}. It captures the change in band energy arising from the interplay of magnetization rotation and SOC. Mathematically, 
we have $\Gamma^{-}_{pq}(\textbf{k}) = \langle{p, \textbf{k}|[\sigma^{-}, H_{so}]|q, \textbf{k}}\rangle$ 
with $\sigma^{-} = \sigma_{x}-i\sigma_{y}$ and 
$\sigma_{x}$, $\sigma_{y}$ are the Pauli matrices. 
$H_{so}$ is the spin-orbit part of the Hamiltonian. The momentum independent $\Gamma_{}^{-}$ in Eq.~\eqref{eqn:dependence_of_alpha} is 
the average of $\Gamma^{-}_{pp}(\textbf{k})$ matrices over the states at 
Fermi energy \cite{Li}. It is given by $\Gamma_{}^{-} = 
\langle{[\sigma^{-}, H_{so}]}\rangle_{E=E_{F}}$. 
We calculate $|\Gamma^{-}|^{2}/|\Gamma^{-}|^{2}_{max}$, and present its angular variation in Fig.~\ref{fig.1}e. Our results highlight that the SOC effect has a strong anisotropy of $\sim$ 590\%, making it the most significant factor giving rise to the large spin-damping anisotropy in our zero temperature calculations. 
In our calculations, F5GT's intrinsic electronic properties give rise to its magnetic anisotropy energy which reflects in the giant Gilbert damping anisotropy. This is in contrast to 
the damping anisotropy caused by interfacial SOC in some materials disappears as the magnetic layer's thickness increases \cite{Chen2018}.
Note that these calculations rely on a finite magnetization value maximum at $0$ temperature. As we approach the critical temperature and the magnetization tends to vanish, the magnetic anisotropy energy and the anisotropy in Gilbert damping will likely vanish. 

{\it Experimental methods.---} Single crystals of F5GT are prepared using the chemical vapour transport method. The average dimension of the single crystals is close to $1$ mm. Energy-dispersive X-ray spectroscopy (EDS) is used to confirm the stoichiometric ratio of the system. X-ray diffraction (XRD) measurement is carried out using a PANalytical X'Pert diffractometer on the bulk single crystals to confirm the good crystallinity of the samples. See Sec. II of the SM~\cite{SM} for the details. 

\begin{figure}
\includegraphics[width=\linewidth]{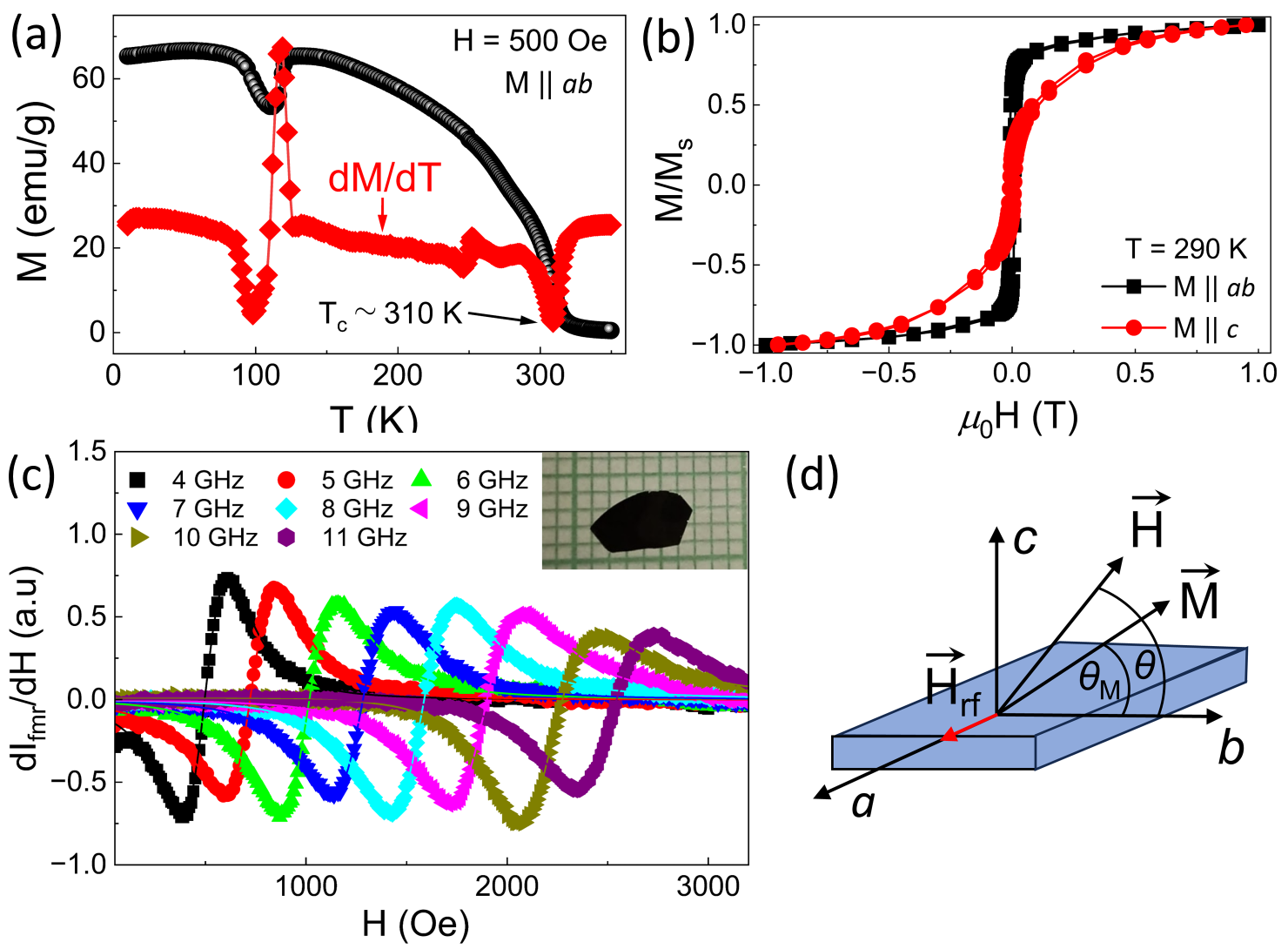}
\caption{(a) Variation of field cooled magnetization with temperature at H $= 500$ Oe applied along the $ab$ plane. (b) Magnetization hysteresis loops at T $= 290$ K for magnetic field applied along the crystal $ab$ plane and $c$ axis. (c) FMR dispersion spectra of a bulk single crystal for external DC magnetic field applied along the $ab$ plane with varying frequencies of the MW field. Inset: Optical image of a bulk F5GT single crystal. (d) Sample geometry considered for energy minimization calculation.}
\label{fig2}
\end{figure}

The temperature dependence of magnetization and the isothermal magnetization hysteresis measurements are carried out using a Quantum Design PPMS. The dependence of in-plane ($ab$) magnetization on the field-cooled (FC) temperature is shown in Fig.~\ref{fig2}a. The Curie temperature, T$_\mathrm{c} = 310 \pm 0.95$ K as derived from the derivative of the magnetization curve. A magnetic reorientation transition is also observed at low temperature. The magnetic hysteresis at $290$ K for magnetic field direction in the $ab$ plane and along the $c$ axis (see Fig.~\ref{fig2}b) show easy in-plane anisotropy. There is no uniaxial anisotropy along the $ab$ plane. The magnetization in the $ab$ plane is saturated at $57$ mT, while along the $c$ axis, it's about 1 T.

We perform FMR measurement using PhaseFMR (NanoSc, Sweden) on the FGT single crystal in the presence of microwave (MW) field of different frequencies at room temperature. The sample is placed on a coplanar waveguide on which an external DC magnetic field is applied in various directions in the $ab$ (in-plane) and along the $c$ axis (out-of-plane) in addition to a MW field along the $ab$ plane, as shown in Fig.~\ref{fig2}d. The polar angle between the applied DC magnetic field and the $ab$ plane is labelled as $\theta$, while the azimuthal angle in the $ab$ plane is labelled as $\phi$. The range of MW frequencies is chosen such that the sample's magnetisation is fully saturated at FMR for all angles ($\theta, \phi$).
 
{\it In-plane magnetization dynamics.---}     
We rotate the bulk single crystal both in-plane ($\phi$) and out-of-plane ($\theta$), with respect to the external static magnetic field direction, to study the orientation-dependent variation in magnetization dynamics. We also did this experiment with different MW frequencies. 
% The $ab$ plane of the bulk single crystal is subjected to in-plane ($\phi$) and out-of-plane ($\theta$) rotations with respect to the external static magnetic field direction, at different fixed MW frequencies to investigate the orientation-dependent variation in magnetization dynamics. 
A few representative spectra at different frequencies at a fixed $\phi$ are shown in Fig.~\ref{fig2}c for the in-plane geometry. After subtracting the offset and the linear $\mathrm{H}$ contribution, the spectrum is fitted with the sum of the symmetric and anti-symmetric Lorentzian function given by, 
%\begin{multline}
%\mathrm{\frac{dI_{FMR}}{dH} = 4A\frac{\Delta H(H-H_{R})}{(4(H-H_{R})^2+(\Delta H)^2)^2}}\\
%\mathrm{-S\frac{\Delta H^2-4(H-H_{R})^2}{(4(H-H_{R})^2+\Delta H^2)^2}}
%\end{multline}
\begin{equation}
\mathrm{\frac{dI_{FMR}}{dH}} = \frac{4\mathrm{A\Delta H(H-H_{R})} - \mathrm{S\left[\Delta H^2-4(H-H_{R})^2\right]}}{\mathrm{\left[4(H-H_{R})^2+\Delta H^2\right]^2}}~.
\end{equation}
Here, $\mathrm{H}$ is the applied field, $\mathrm{\Delta H}$ is the line-width, $\mathrm{H_{R}}$ is the resonant field. $\mathrm{A}$ and $\mathrm{S}$ are the weights of the anti-symmetric and symmetric Lorentzian functions, respectively. We obtain parameters such as $\mathrm{H_R}$ and $\mathrm{\Delta H}$ from the Lorentzian fit at various MW frequencies, $\mathrm{f}$ for further analysis.

For systems exhibiting significant out-of-plane hard axis anisotropy and no uniaxial anisotropy in the easy plane, the free energy density of the magnetic configuration can be expressed as
\begin{eqnarray}
\mathrm{F}&=& \mathrm{-MH\,\left[\cos\theta \cos\theta_M \cos(\phi-\phi_M)+\sin\theta \sin\theta_M\right]}\nonumber \\
& & +\mathrm{K\cos^2\theta_M-2\pi M^2 \cos^2 \theta_M}~.
\end{eqnarray}
Here, $\mathrm{\theta_M}$ and $\mathrm{\phi_M}$ represent the polar and azimuthal angles corresponding to the magnetization vector, respectively. $\mathrm{M_{eff}}$ denotes the effective magnetization of the system, which is linked to the saturation magnetization $\mathrm{M_s}$ and the perpendicular anisotropy constant $\mathrm{K}$ through the equation $\mathrm{4\pi M_{eff}=4\pi M_s-(2K/M_s)}$ in CGS system.

Using energy minimization, we derive the relationship between the applied magnetic field, $\mathrm{H_{R}}$ and the MW frequency, $\mathrm{f}$, known as the Kittel equation~\cite{CGT}.  
\begin{equation}
\mathrm{\left(\frac{2\pi f}{\gamma}\right)^2=H_1\times H_2}~,
\label{Kittel}
\end{equation}
where 
\begin{equation}
\mathrm{H_1=H_{R}\cos(\theta-\theta_{M})-4\pi M_{eff}sin^2\theta_{M}}~,
\end{equation}
and,
\begin{equation}
\mathrm{H_2=H_{R}\cos(\theta-\theta_{M})+4\pi M_{eff}\cos2\theta_{M}}~.
\end{equation}
Here, $\gamma$ is the gyromagnetic ratio. The dimensionless form of $\gamma$ is called Landé g-factor, given by $\mathrm{g=\hbar \gamma/\mu_B}$, where $\mathrm{\hbar}$ and $\mathrm{\mu_B}$ are reduced Planck constant and Bohr magneton, respectively.
\begin{figure*}
\includegraphics[width=\linewidth]{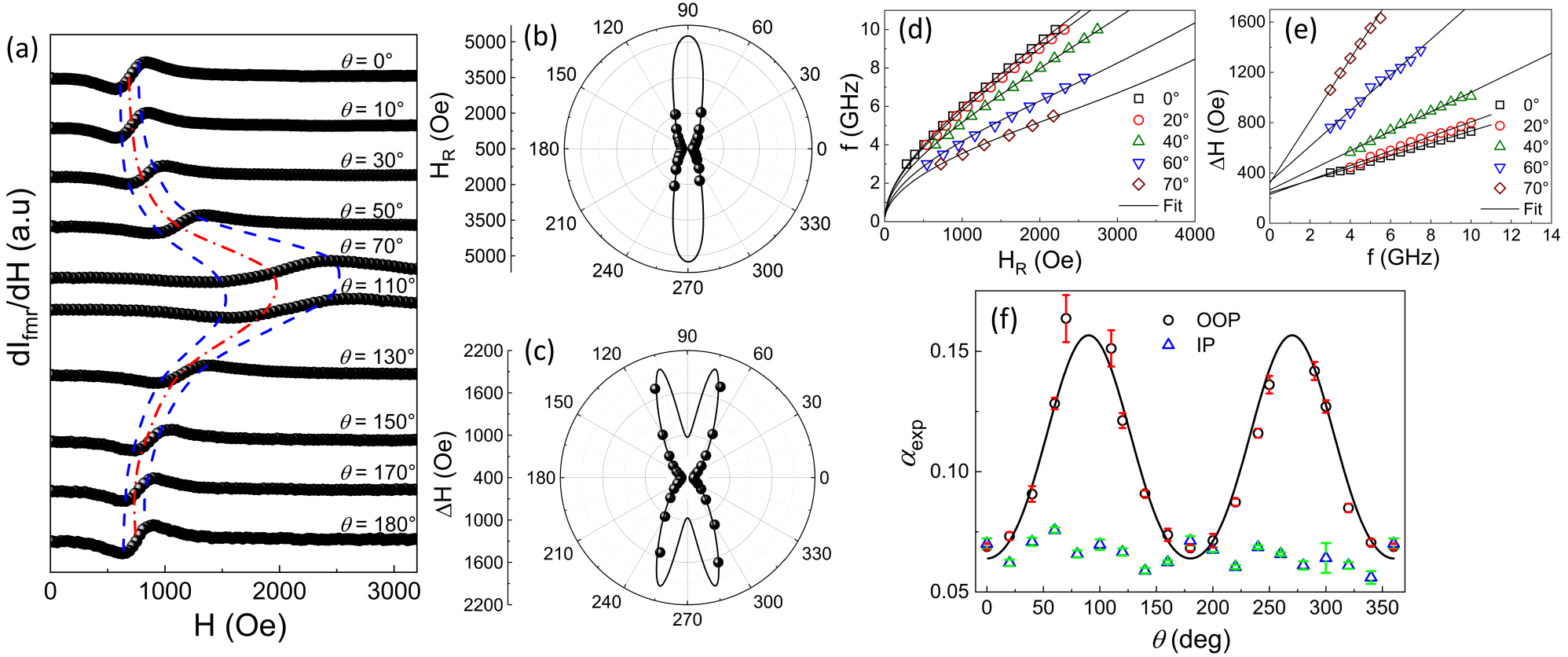}
\caption{(a) Evolution of the FMR spectra with polar angle at a constant MW frequency, f $= 5$ GHz and temperature, T $= 290$ K. Red and blue lines track the change in the resonant field and linewidth, respectively. Out-of-plane two-fold anisotropy can be seen in (b) resonant magnetic field and (c) line-width broadening in bulk single crystal. Variation of (d) frequency with resonant field $\mathrm{H_R}$, and (e) linewidth $\mathrm{\Delta H}$ with f at different $\theta$. Open symbols represent measured data points, and solid black lines show fitted curves. (f) In-plane isotropy (symbol labelled as IP) and out-of-plane (symbol labelled as OOP) two-fold anisotropy in Gilbert damping factor $\mathrm{\alpha}$. The black solid line shows the fit to Eq.~\ref{eddy_fit}.
\label{fig3}}
\end{figure*}
The angles $\mathrm{\theta}$ and $\mathrm{\theta_M}$ are related by,
\begin{equation}
\mathrm{sin2\theta_{M}=\frac{2H_R}{4\pi M_{eff}}\sin(\theta-\theta_{M})}~.
\end{equation}
This shows if $\mathrm{\theta=0^\circ}$, i.e. the external DC magnetic field is along the $ab$ plane, then $\mathrm{\theta_M}$ is also $\mathrm{0^\circ}$. Thus, for in-plane orientation of the magnetic field, Eq.~\ref{Kittel} simplifies to
\begin{equation}
\mathrm{f=\frac{\gamma}{2\pi}\sqrt{H_{R}(H_{R}+4\pi M_{eff})}}~.
\label{gfac1}
\end{equation}
We obtain $\mathrm{g=2.061}$ and $\mathrm{4\pi M_{eff}=3340}$ Oe, irrespective of the value of $\phi$. The small deviation of the g-factor from $\mathrm{g = 2}$ may be attributed to an orbital contribution to the magnetization due to spin-orbit coupling~\cite{Alahmed}.

As discussed earlier, the line-width broadening $\mathrm{\Delta H}$ arise from a combination of intrinsic contributions from Gilbert damping and extrinsic contributions from sample imperfections, eddy current damping, scattering phenomena, etc. Usually, $\mathrm{\Delta H}$ demonstrates a linear trend with variation in frequency and can be written as,
\begin{equation}
\mathrm{\Delta H=\frac{4\pi \alpha_{exp}}{\gamma\Omega}f+\Delta H_0}~,
\label{eq6}
\end{equation}
where $\mathrm{\alpha_{exp}} $ is the effective damping factor, which can be calculated experimentally from FMR spectra,  $\mathrm{\Delta H_0}$ is in-homogeneous line-width broadening. The quantity $\mathrm{\Omega}$ is frequently referred to as the dragging function, arising from the drag effect, i.e. misalignment between the magnetic field and the magnetization of the system, and is given by,
\begin{equation}
\mathrm{\Omega=\cos(\theta-\theta_{M})-H_R\frac{3H_1+H_2}{H_2(H_1+H_2)}\sin^2(\theta-\theta_{M})}~.
\label{eq7}
\end{equation}
For in-plane configuration, $\mathrm{\theta=\theta_M=0^\circ}$ and Eq.~\ref{eq6} reduces to 
\begin{equation}
\mathrm{\Delta H=\frac{4\pi \alpha_{exp}}{\gamma}f+\Delta H_0}~.
\label{eq8}
\end{equation}
Damping parameter $\mathrm{\alpha_{exp}}$ for in-plane orientations are determined by fitting the $\mathrm{\Delta H}$ vs. $\mathrm{f}$ data to Eq.~\ref{eq8} at different $\phi$ angles. After fitting the in-plane FMR spectra with this equation, we determine $\mathrm{\alpha_{exp}=0.0603}$ and $\mathrm{\Delta H_0=250.23}$ Oe 
% for the in-plane configuration
, which is independent of $\phi$, suggesting the absence of uniaxial anisotropy in the $ab$ plane.

{\it Out-of-plane magnetization dynamics.---}
Although no uniaxial anisotropy is observed in $\mathrm{H_R}$ in the $ab$ plane, the plot of $\mathrm{\theta}$ dependent resonant magnetic field indicates the presence of two-fold anisotropy in the $\hat{\theta}$ direction. The easy plane aligns with the $ab$ plane, while the hard axis is oriented along the c-axis. The variation of the resonant magnetic field with polar angle $\theta$ can be accurately described by the model presented in Eq.~\ref{Kittel}, demonstrating an excellent fit with the experimental data shown in Figure~\ref{fig3}b. Furthermore, by fitting Eq.~\ref{Kittel} to the f vs $\mathrm{H_R}$ data taken at various polar angles $\theta$, we determine the angular variation of the gyromagnetic ratio $\gamma$ and the Landé g-factor. We shall come back to this point later. 

To extract $\mathrm{\alpha_{exp}}$ at various $\theta$ values, we prepare two sets of data: $\mathrm{\Delta H}$ vs. $\theta$ at a fixed $\mathrm{f=5}$ GHz and $\mathrm{\Delta H}$ vs. $\mathrm{f}$ at various $\theta$ values. Both datasets were fitted with Eq.~\ref{eq6} in order to determine the $\mathrm{\alpha_{exp}}$ values corresponding to different orientations $\theta$. Although $\mathrm{\Delta H}$ and $\mathrm{\alpha_{exp}}$ is isotropic against $\phi$ rotation as discussed in the previous section, there is clear two-fold anisotropy in $\mathrm{\Delta H}$ with $\theta$ variation (Fig.~\ref{fig3}c). Besides, our investigations reveal that the inhomogeneous linewidth broadening exhibits minimal out-of-plane anisotropy (see the supplementary material~\cite{SM} for detail). Curiously, $\mathrm{\alpha_{exp}}$ is found to show a two-fold anisotropy with the polar angle variation (Fig.~\ref{fig3}f) with a maximum of almost 300$\%$ increase compared to the minimum damping at $\theta=0$ at room temperature. Such large variation in $\mathrm{\alpha_{exp}}$ with variation in $\theta$ is not only rare but theoretically non-trivial as the anisotropy in $\mathrm{\Delta H}$ does not necessarily guarantee anisotropy in $\mathrm{\alpha_{exp}}$. Usually, the dragging function $\Omega$ is $\theta$ independent and maintains a constant value of 1. In such a scenario, Eq.~\ref{eq6} reduces to Eq.~\ref{eq8}. However, the existence of anisotropy within a system, stemming from factors like the demagnetizing field, perpendicular magnetic anisotropy (PMA), surface anisotropy, etc., can give rise to a dragging phenomenon as the magnetization vector either lags or leads the magnetic field orientation, resulting in a pronounced $\theta$ dependence of the dragging function $\Omega$ itself. This phenomenon may generate an anisotropic $\mathrm{\Delta H}$, even though $\mathrm{\alpha_{exp}}$ remains constant. Moreover, the angular dependency of two magnon scattering can also contribute to the angular variation of $\mathrm{\Delta H}$, not necessarily leading to the angular variation of $\mathrm{\alpha_{exp}}$ (see supplementary material~\cite{SM} for details).  

\begin{figure}
\includegraphics[width=0.75\linewidth]{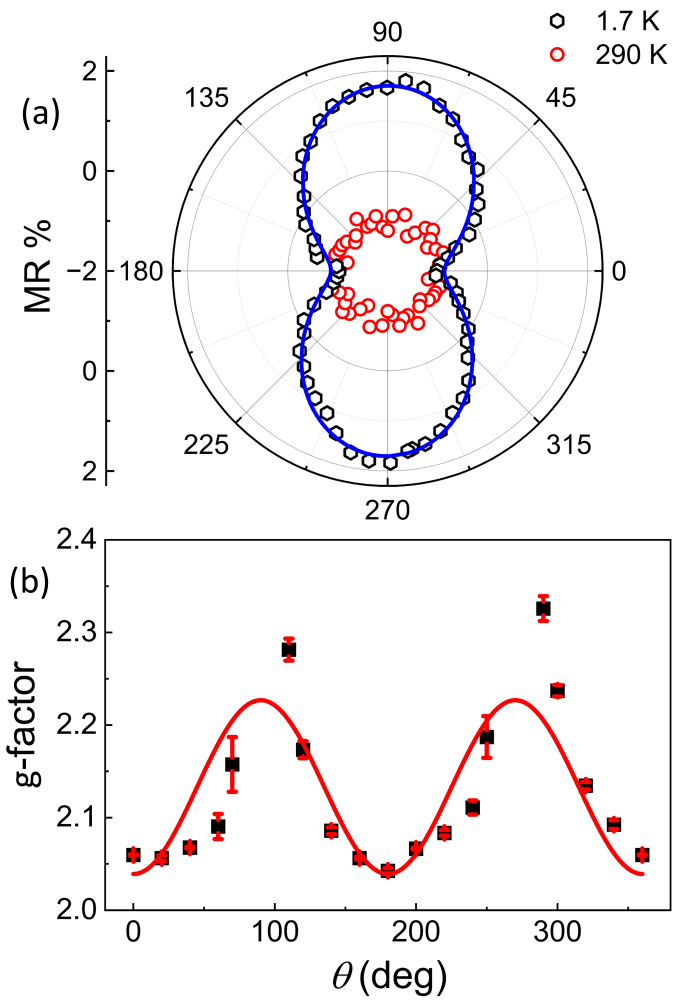}
\caption{(a) The angular variation of the magneto-resistance at 9 T magnetic field at temperatures of 1.7 K, and 290 K. The angular dependence of the low-temperature magneto-resistance is similar to the angular dependence of the magnetic anisotropy energy or the Gilbert damping in Fig.~\ref{fig.1} with a $\cos^2\theta$ dependence (solid blue line shows the fit). In contrast, the high-temperature magneto-resistance curve has negligible anisotropy. (b) The out-of-plane anisotropy in Landé g-factor, measured at 290 K. The continuous (red) line is the fit of the experimental data using Eq.~\ref{Nagata}. 
\label{fig4}}
\end{figure}

We are left with two possibilities: 1) that the polar angle dependence is intrinsic and arises due to spin-orbit interaction as already predicted theoretically earlier; 2) that the polar angle dependence is extrinsic and arises due to the MW field-induced eddy current contributions. 

The damping factor $\alpha_{\mathrm{ed}}$ due to eddy current (in SI unit) is given by~\cite{Lock, Bailey} 
\begin{equation}
\mathrm{\alpha_{ed}=\frac{\mu_0^2 \gamma M_s t^2} {12 \rho}}~,
\label{eddy}
\end{equation}
where $\mathrm{\mu_0}$ is the vacuum permeability, M$\mathrm{_s}$ is the saturation magnetization, and $\mathrm{t}$ and $\rho$ are the thickness and resistivity, respectively, of the sample. Skin-depth of any metallic sample is given by $\mathrm{\delta\approx\sqrt{2\rho/(\omega \mu_0 \mu_r)}.}$ For sample thickness higher than the skin-depth, the current density is localized primarily along the sample edge. Due to this localized distribution, the induced current is approximated to the first order as a current loop along the sample edge.

For a sample of skin depth $\delta$, thickness t subjected to a static field applied along the $ab$ plane, we modify the expression in Eq.~\ref{eddy} for eddy current damping as follows.
\begin{equation}
\mathrm{\alpha_{ed}=\frac{1}{3\rho} \mu_0^2 \gamma M_s \frac{(t/2)^3-(t/2-\delta)^3}{t/2}}~.
\end{equation}
The $\theta$ dependence of the damping factor due to the eddy current is then given by (see Sec. V of the SM~\cite{SM} for details of the derivation),
\begin{equation}
\mathrm{\alpha_{ed}(\theta)} = \frac{\mu_0^2 \gamma {\mathrm{M_s}} a^2}{6\rho} (1+\sin^2\theta)\,\sin[\Theta_{ab}
 +(\Theta_c-\Theta_{ab})\sin^2\theta]~.
\end{equation}
Here, $a=\mathrm{\sqrt{\left[(t/2)^3-(t/2-\delta)^3\right]/(t/2)}}$ is the effective length scale due to skin effect, $\mathrm{(1+\sin^2\theta)/2}$ is the correction for the eddy current redistribution due to $\theta$ rotation of magnetization and $\sin[\Theta_{ab}+(\Theta_c-\Theta_{ab}) \sin^2\theta]$ is cone angle correction, where $\Theta_{ab}$ and $\Theta_{c}$ are cone angles for $\mathrm{H}\parallel ab$ and $\mathrm{H}\parallel c$, respectively. The experimentally observed damping factor is thus given by,
\begin{equation}
\mathrm{\alpha_{\mathrm{exp}}(\theta)=\alpha +\alpha_{ed}(\theta)}~,
\label{eddy_fit}
\end{equation}
where $\mathrm{\alpha}$ is the intrinsic Gilbert damping factor.

Using the system parameters $\rho = 92.10 \,\mu \Omega .\mathrm{cm}$, $\mathrm{M_s} = 0.36 \,\mathrm{T}$ and 
$\mu_\mathrm{r} = 102.06$, the skin-depth for our system turns out to be  $\delta = 0.504$ $\mu \mathrm{m}$. We take typical cone angle values of $\Theta_{ab} = 3^\circ$ and $\Theta_{c} = 10^\circ$. It is found that this analysis is immune to a slight variation in these values (see Sec. V of the SM~\cite{SM} for details). We get $\alpha=0.048$ and $\mathrm{t=28.46}$ $\mathrm{\mu m}$ by fitting Eq.~\ref{eddy_fit} to the experimental data (see Fig.~\ref{fig3}f), keeping $\mathrm{t}$ and $\alpha$ as free parameters. These values are quite close to experimentally found values of $\alpha_{\mathrm{exp}} \sim 0.06$ for $\theta=0^\circ$ and $\mathrm{t\sim 30}$ $\mathrm{\mu m}$. This indicates that eddy current-induced damping is mostly responsible for the observed anisotropy in the damping factor. Near room temperature, spin-orbit coupling contribution is largely suppressed. This diminished effective spin-orbit coupling strength near room temperature is also reflected in the weak polar angular dependency of magneto-resistance (MR) at 290 K as shown in Fig.~\ref{fig4}a (see data in red colour). In contrast, the MR at 1.7 K in Fig.~\ref{fig4}a (see data in black with blue fit) is highly anisotropic with a fit of the form, $\rho = \rho_c - \Delta \rho \cos^2 \theta$. This $\cos^2\theta$ anisotropy is similar to the calculated anisotropy at zero temperature in the MAE and in the Gilbert damping factor presented in Fig.~\ref{fig.1}b and c. Thus, the predicted intrinsic Gilbert damping anisotropy due to spin-orbit interaction might show up at low temperatures, given that the eddy current contribution is likely to increase only marginally.

Fig.~\ref{fig4}b depicts the angle-dependent out-of-plane g-factor as found in the F5GT system. This angular variation in the g-factor usually appears due to presence of anisotropic spin-orbit coupling. Beyond the spin-orbit interaction, critical fluctuations and anisotropic exchange interactions, as demonstrated by Nagata et al.~\cite{Nagata1, Nagata2, Nagata3, Nagata4}, can also contribute to the experimental observation of the angle-dependent g-factor. With the magneto-resistance data demonstrating the negligible impact of spin-orbit contribution at room temperature, we can reasonably deduce that these critical dynamics and anisotropic exchange induced g-shift are the predominant mechanisms behind the anisotropic behavior of the g-factor in F5GT, the experimental temperature (290 K) being close to its critical temperature (310 K). 

The Nagata theory, extended to the ferromagnetic state, gives the expression~\cite{NagataTheory} of the anisotropic g-factor as:
\begin{equation}
\mathrm{g(\theta)=\Delta g(3\sin^2 \theta-1)+g_{iso}}~, 
\label{Nagata}
\end{equation}
with $\mathrm{g_{iso}}$ being the isotropic component of the g-factor and $\mathrm{\Delta g}$ is the g-shift amplitude due to critical fluctuations and anisotropic exchange interactions. However, the anisotropic exchange is anticipated to be small enough to make a negligible contribution to the g-shift amplitude of F5GT, similar to CrGeTe$_3$ (CGT)~\cite{NagataTheory}. This is reasonable as CGT and F5GT belong to comparable universality classes. While CGT fits into quasi-isotropic Heisenberg spin models~\cite{CGTModel}, F5GT lies between the 3D Heisenberg model and the 3D XY model~\cite{FGTModel}. 

As seen from Fig.~\ref{fig4}b, the extracted g-factor fits well with the angular dependence proposed by this theory. From the fit, we obtain the isotropic component $\mathrm{g_{iso}=2.101}$, which is slightly higher than the spin-only value of $\mathrm{g=2}$, suggesting a finite but marginal orbital contribution to the magnetization of F5GT even at room temperature. The g-shift amplitude for F5GT turns out to be $\mathrm{\Delta g=0.063}$, which is also the g-shift along the $ab$ plane ($\mathrm{g}_{ab}$). Along the $c$ axis, the g-shift is $\mathrm{g}_c=2\mathrm{\Delta g=0.126}$, slightly smaller compared to that of CGT ($\Delta \mathrm{g}_c [\mathrm{70\,K}]\sim0.20$)~\cite{NagataTheory}. This is probably attributed to the fact that the measurement is not being performed exactly at the Curie temperature of F5GT. Additionally, in the absence of strong anisotropic exchange interaction, the g-shift is confined to a smaller temperature window. This weak anisotropic exchange is also the reason behind the smaller g-shift in both F5GT and CGT when compared to that of CrSiTe$_3$ ($\Delta \mathrm{g}_c [\mathrm{35\,K}]\sim0.70$), which exhibits significant anisotropy in exchange interaction owing to the quasi-2D Ising model that it fits into ~\cite{NagataTheory}. From another perspective, this offers an alternative method to probe anisotropic exchange interactions in ferromagnetic systems, which is crucial for understanding the nature of magnetism as it approaches the 2D limit.

In summary, our findings strongly indicate that the out-of-plane anisotropic linewidth broadening in Fe$_5$GeTe$_2$ (F5GT) at room temperature is primarily due to the eddy current contribution rather than spin-orbit interaction. Although theoretical calculations suggest an intrinsic origin of anisotropy due to spin-orbit interaction at low temperatures, it does not contribute to the anisotropy of the Gilbert damping factor at room temperature, nor does it play a significant role in the observed g-factor anisotropy. We demonstrate that the intrinsic polar angular variation of the g-factor is mainly due to enhanced anisotropic critical spin fluctuations, given that the critical temperature of F5GT is close to room temperature. Our findings offer new insights into the fundamental physics of magnetization dynamics in van der Waals magnets. Additionally, the observed decoupling between magneto-crystalline anisotropy and magnetization damping can lead to exciting possibilities for future applications in spintronics.

{\it Acknowledgements.---} SM and AA acknowledge Department of Science and Technology, India, DST Nanomission,  for financial support. AB thanks PMRF for financial support. We acknowledge the high-performance computing facility at IIT Kanpur for computational support. NJ thanks Atasi Chakraborty for the discussions.
% We thank the Param Sanganak high performence computing facilities at IIT Kanpur for computational support. 

\bibliography{F5GT_references}

\end{document}